# Embroidered Antenna Characterization for Passive UHF RFID Tags

Philip H. Gordon*, Rex Chen*, Huiju Park[+] and Edwin C. Kan*
*Electrical and Computer Engineering, Cornell University, Ithaca, NY, USA, phg35@cornell.edu
[+]Department of Fiber Science and Apparel Design, Cornell University, Ithaca, NY, USA

## Summary

For smart clothing integration with the wireless system based on radio frequency (RF) backscattering, we demonstrate an ultra-high frequency (UHF) antenna [1] constructed from embroidered conductive threads [2, 3]. Sewn into a fabric backing, the T-match antenna design mimics a commercial UHF RFID tag, which was also used for comparative testing. Bonded to the fabric antenna is the integrated circuit chip dissected from another commercial RFID tag, which allows for testing the tags under normal EPC Gen 2 operating conditions. We find that, despite of the high resistive loss of the antenna and inexact impedance matching [4], the fabric antenna works reasonably well as a UHF antenna both in standalone RFID testing, and during variety of ways of wearing under sweaters or as wristbands. The embroidering pattern does not affect much the feel and comfort from either side of the fabrics by our sewing method.

## Motivation

To enable wireless sensing and data transmission of useful wearer conditions, integration of battery-free sensor tags unto clothing is technically feasible and affordable. Due to wearing and laundry considerations, passive UHF RFID tags such as those employed in item-level inventory monitoring offer many advantages in garment manufacturing and care. Conventional RFID tags are often built on a paper or plastic substrate with laminated plastic covering an aluminum antenna and a small integrated-circuit (IC) chip. This packaging is cost effective for logistics-controlled tags, but does not integrate well for wear comfort and for laundry durability. Printed aluminum antennas and the non-fabric substrate cause additional limits in normal wear, aesthetic designs, and washing/drying options. We minimize the amount of non-fabric materials in the tag by constructing the antenna using conductive threads, embroidered directly onto the fabric [1]. Conductive epoxy was chosen as the interconnect method to the IC chip to minimize the impact on read distance over the course of wear and wash cycles [2].

## Results

Conductive threads were used to embroider a UHF T-match antenna pattern onto a piece of polyester fabric (Fig. 1). The conductive thread has a single-thread resistivity of .7Ω/cm, with strand-to-strand resistance dependent on the tightness of the stitching. Double-sided stitching was used to decrease the resistance, and the total DC resistance across the antenna was around 51Ω. While dense stitching has been reported to cause unexpected losses [3], this was deemed necessary to counteract the contact resistance to the IC chip, measured at 47Ω. An RFID tag IC chip was cut from a commercial tag, and connected to the antenna via conductive epoxy after removing the initial plastic passivation and adhesive layers with a two-hour toluene submersion. The fabric tag was then tested using an Impinj Speedway® R420 reader with an output power of 32.5dBm and a Laird S9028PCR panel antenna, side by side with a commercial UHF EPC paper-based tag. The two tags performed very similarly out to 2.5m (Fig. 2a) with typical oscillation of received signal strength indicator (RSSI) for indoors environment. To study the perspiration influence, 100mM saline solution was used to simulate sweat. Partial saline wetting decreased RSSI for 6 – 10dB probably due to mismatched impedance (Fig. 2b), but the embroidered tag remains functional at a reduced distance. In addition, the fabric tag was tested on the chest [4], under a coat on the chest, and on the wrist as a wristband. While significant attenuation was noted (specifically with the wristband configuration, which was only readable out to 30cm due to the reduced radar cross section), the fabric tag did function in all wearable cases (Table 1) with EPC protocols.

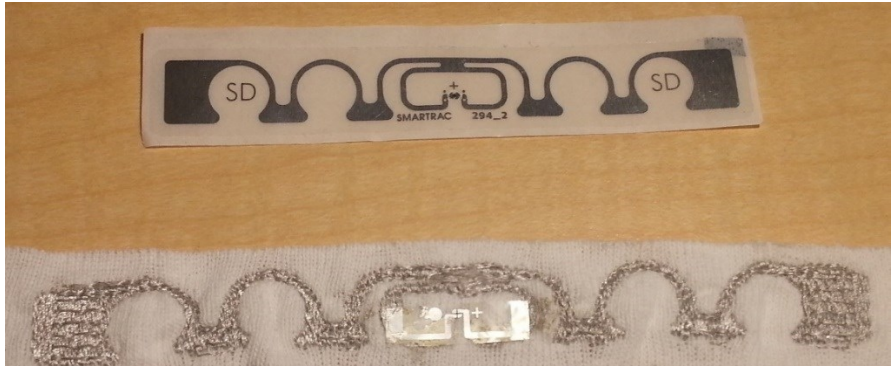

**Fig. 1:** Photograph of commercial Smartrac® ShortDipole UHF RFID tag (top) next to the fabric based antenna with an IC chip connected in the center (bottom). The fabric antenna, while 15% larger and 300x higher resistance, still functioned similarly out to 2.5m.

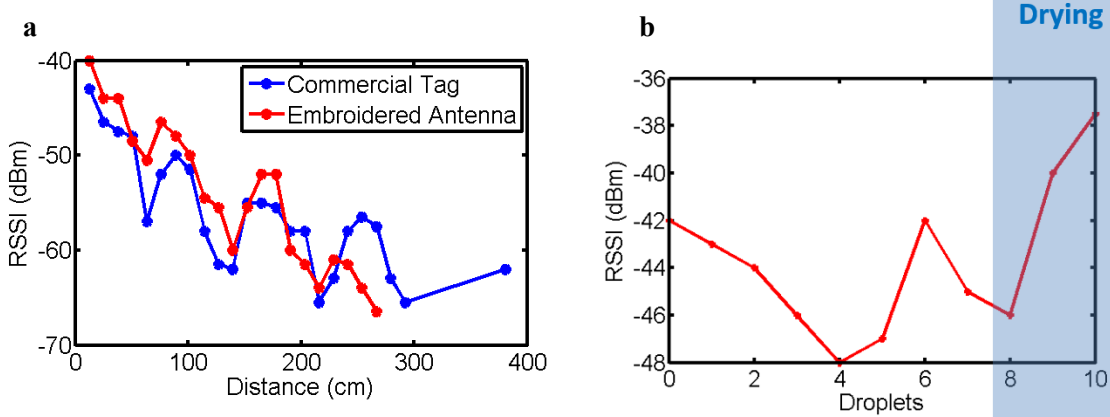

**Fig. 2:** a) Received signal strength indicator (RSSI) vs. reading distance for both the embroidered fabric tag (red) and the commercial control tag (blue). Comparable performance was achieved out to 2.5m, before signal was lost for the embroidered antenna. b) Embroidered antenna RSSI response to 100mM NaCl saline solution. Droplets were added until full saturation at droplet 7, after which recordings were taken until fully dry at marker 10. Partial wetness decreased signal strength, which had worse performance than full saline saturation probably due to impedance mismatch. Returning to a dry state showed no lasting effects.

| Distance (cm) | RSSI On Shirt (dBm) | RSSI Under Sweater (dBm) |
|---|---|---|
| 30 | −44 | −59 |
| 90 | −49 | −64 |
| 150 | −55.5 | −64.5 |

**Table 1:** RSSI at 60cm intervals for the tag worn on the front of a shirt, with either clear line of sight to the reader, or with a sweater covering the tag. Wristband configuration can only be read from 30cm.